\pdfoutput=1

\documentclass[11pt, reqno,preprint]{article}
\usepackage{jheppub}
\usepackage{epsfig}
\usepackage{amssymb}
\usepackage{amsmath}
\usepackage{mathrsfs}
\usepackage{hyperref}
\usepackage{multirow}
\usepackage{dsfont}

\def\be{\begin{equation}}
\def\ee{\end{equation}}
\def\ba{\begin{eqnarray}}
\def\ea{\end{eqnarray}}
\def\nl{\nonumber\\}

\def\CP1{\mathbb{CP}^1}
\def\SL2C{\mathrm{SL}(2,\mathbb{C})}

\def\Z2{\mathbb{Z}_2}

\title{Scattering Equations and KLT Orthogonality}
\author{Freddy Cachazo${}^{a}$, Song He${}^{a,b}$ and Ellis Ye Yuan${}^{a,c}$}
\affiliation[a]{Perimeter Institute for Theoretical Physics, Waterloo, ON N2L 2Y5,
Canada}
\affiliation[b]{School of Natural Sciences,
Institute for Advanced Study, Princeton, NJ 08540, USA}
\affiliation[c]{Physics Department, University of Waterloo, Waterloo, ON N2L 2Y5,
Canada}
\emailAdd{fcachazo, she, yyuan@perimeterinstitute.ca}

\abstract{Several recent developments point to the fact that rational maps from $n$-punctured spheres to the null cone of $D$ dimensional momentum space provide a natural language for describing the scattering of massless particles in $D$ dimensions. In this note we identify and study equations relating the kinematic invariants $s_{ab}$ and the puncture locations $\sigma_c$, which we call the {\it scattering equations}. We provide an inductive algorithm in the number of particles for their solutions and prove a remarkable property which we call {\it KLT Orthogonality}. In a nutshell, KLT orthogonality means that ``Parke-Taylor" vectors constructed from the solutions to the scattering equations are mutually orthogonal with respect to the Kawai-Lewellen-Tye (KLT) bilinear form. We end with comments on possible connections to gauge theory and gravity amplitudes in any dimension and to the high-energy limit of string theory amplitudes.
}

\begin{document}
\maketitle

\section{Introduction}

Recently the tree-level S-matrix of a variety of theories has been found as an integral over the moduli space of maps from the $n$-punctured sphere into the null light cone in momentum space \cite{Witten:2003nn,Roiban:2004yf,Cachazo:2012da,Cachazo:2012kg,Huang:2012vt,Cachazo:2013iaa}. In all cases a given point in the space of kinematic invariants is mapped into $(n-3)!$ points in the moduli space. In this note we point out that the system of polynomial equations that connects the two spaces is universal and independent of the spacetime dimension. In section \ref{sec:eq} we derive the basic set of equations connecting the space of kinematic invariants and that of the puncture locations; we call them the {\it scattering equations}. In section \ref{sec:solve} we give an algorithm for finding all solutions to the scattering equations which proves that the number of solutions is also $(n-3)!$ in any dimension. In order to motivate the fact that the scattering equations are indeed the backbone of massless particle scattering in any dimension, in section \ref{sec:factorization} we study how general kinematic invariants can nicely be extracted and show the behavior of the equations in factorization limits.

The solutions to the scattering equations also satisfy a remarkable property which we call {\it KLT orthogonality}. This was conjectured in the context of four dimensional scattering by one of the authors and Geyer \cite{Cachazo:2012da}. KLT orthogonality states that when the standard field theoretic version of the Kawai-Lewellen-Tye (KLT) construction~\cite{Kawai:1985xq} is interpreted as a bilinear form acting on an $(n-3)!$ dimensional vector space, the vectors made of Parke-Taylor-like functions constructed from the puncture locations evaluated on different solutions are mutually orthogonal with respect to the KLT bilinear. In section \ref{sec:KLT} we make precise the elements entering the conjecture and give a proof of its validity in any dimension.

In four dimensions, KLT orthogonality was crucial for constructing a formula for gravity amplitudes from rational maps starting with the Witten-RSV formula~\cite{Witten:2003nn,Roiban:2004yf} for gauge theory amplitudes \cite{Cachazo:2012da}. It is natural to assume that given the validity of KLT orthogonality in any dimension, such a connection between gauge theory and gravity S-matrices should also exist. We end in section \ref{sec:discussion} with comments on these connections and also to the high energy limit of string scattering amplitudes.

\section{Scattering Equations}~\label{sec:eq}

In this section we derive the scattering equations starting with the connection between the scattering data of $n$ massless particles and maps from the $n$-punctured sphere into the null cone in $D$ dimensional momentum space. Given a set of $n$ null vectors in $D$ dimensions $\{ k_1^\mu ,k_2^\mu ,\ldots ,k_n^\mu \}$ that satisfy momentum conservation, the map is given by~\cite{Cachazo:2013iaa}
\be
k^\mu_a = \frac{1}{2\pi i}\oint_{|z-\sigma_a|=\epsilon} dz \frac{p^\mu(z)}{\prod_{b=1}^n(z-\sigma_b)}~\label{maps}
\ee
where $p^\mu(z)$ is a collection of $D$ degree $n-2$ polynomials.

Given that all $k_a^\mu$ are null vectors it is clear that $p(\sigma_a)^2$ must vanish for all $a$. Since $p(z)^2$ is a polynomial of degree $2n-4$, knowing $n$ roots is not enough to find it. We need $n-3$ additional conditions ($p(z)^2$ is not monic). The extra conditions turn out to have a very elegant origin. They are the conditions needed for the vector $p^\mu(z)$ to be null for all $z$. In other words, $p^\mu(z)$ must be a map from $\mathbb{CP}^1$ to the null cone in (complexified) momentum space. Clearly, if $p^2(z)=0$ then its derivative must also vanish, so $p(z)\cdot p'(z)=0$. It is natural to evaluate this condition on the $n$ puncture locations $\sigma_a$. A simple exercise shows that out of the $n$ conditions
\be
p(\sigma_a)\cdot p'(\sigma_a)=0
\label{pconstraints}\ee
with $a$ in $\{1,2,\ldots ,n\}$, only $n-3$ are linearly independent. These $n-3$ equations are exactly the remaining equations needed to determine $p^\mu(z)$.

Alternatively, the equations (\ref{maps}) can also be treated as linear equations for the coefficients of the polynomial
\be
p^\mu(z) = p_0^\mu+p_1^\mu z+ \cdots + p^\mu_{n-2}z^{n-2}
\ee
whose solution is
\be
p_\alpha^\mu=\sum_{a=1}^n\{\sigma\}_{a}^{n-1-\alpha}k_a^\mu,\quad \alpha\in \{0,1,\ldots,n{-}2\},
\ee
where the symbol $\{\sigma\}_{b}^m$ is defined as the symmetrized product of $m$ $\sigma$'s which do not involve $\sigma_b$
\be
\{\sigma\}_{b}^m=(-1)^m\sum_{\{a_i\}\subseteq\{1,\ldots,n\}\backslash\{b\}}\sigma_{a_1}\sigma_{a_2}\ldots\sigma_{a_m}.
\ee
Plugging this into the constraints \eqref{pconstraints} turns them into
\be
p(\sigma_a)\cdot p'(\sigma_a)
=v_a\sum_{b\neq a}k_a\cdot k_b\sum_{m=1}^{n-2}m\{\sigma\}_b^{n-1-m}\sigma_a^{m-1}=\frac{v_a^2}{2}\sum_{b\neq a}\frac{s_{ab}}{\sigma_a-\sigma_b},
\label{deri}\ee
with $v_a=\prod_{b\neq a}(\sigma_a-\sigma_b)$ and $s_{ab}=2k_a\cdot k_b$ are the standard kinematic invariants\footnote{The last equality in \eqref{deri} is obtained by adding to the second expression a term $v_a \sum_{b\neq a}k_a\cdot k_b (n{-}1)\sigma_a^{n{-}2}$, which vanishes due to momentum conservation.}. Since $v_a$ can never be singular (for generic external data), we can eliminate it from the equations and obtain
\be\label{scatteringequations}
\sum_{b\neq a}\frac{s_{ab}}{\sigma_a-\sigma_b}=0,\quad a\in\{1,\ldots,n\}.
\ee
We call these the \emph{scattering equations}, which are the main object of study in this paper.

\section{Solving the Scattering Equations}~\label{sec:solve}

The scattering equations look deceivingly simple but once the denominators are removed they turn out to be an intricate system of polynomial equations in $n-3$ variables which can resist straightforward numerical algorithms. In \cite{Cachazo:2013iaa} it was proven that for four dimensional kinematics the total number of solutions is $(n-3)!$. In this section we provide a simple numerical algorithm for finding all solutions which makes it clear why the number of solutions is also $(n-3)!$ in any number of dimensions. The algorithm is inspired by the behavior of the equations in soft limits.

Given that only $n-3$ equations are linearly independent we select
\be
\sum_{b\neq a}\frac{s_{ab}}{\sigma_a-\sigma_b}=0,\quad a\in\{4,5,\ldots,n\}.\label{sol}
\ee
and use $\SL2C$ invariance\footnote{To see the $\SL2C$ invariance of the equations one has to use momentum conservation.} to fix the value of $\sigma_1\to \infty$, $\sigma_2=0$ and $\sigma_3=1$.

The algorithm for finding solutions is inductive in nature: Introduce a parameter $\epsilon$ which takes values in the interval $[0,1]$ and define
\be
s_{nb}(\epsilon ) =\epsilon s_{nb}, \quad b\in \{1,2,\ldots ,n-1\}.
\ee
Replacing $s_{nb}$ with $s_{nb}(\epsilon )$ in the original equations (\ref{sol}) gives a set of $\epsilon$-dependent equations.

The new equations evaluated at $\epsilon = 0$ are
\be
\sum_{b=2,b\neq a}^{n-1}\frac{s_{ab}}{\sigma_a-\sigma_b}=0,\quad a\in\{4,5,\ldots,n-1\}.
\ee
Note that the sum starts at $b=2$ due to our gauge choice.

The last equation in (\ref{sol}), i.e, for $a=n$, drops out and the new system is exactly of the form of the equations for $n-1$ particles. At first sight, one would think that the new equations cannot be interpreted exactly as physical equations since the kinematic invariants entering them are not in the kinematic space for $n-1$ particles; they are still the original $n$-particle ones. However, since none of the invariants containing $k_1^\mu$ appear in the equations (recall that $\sigma_1 \to \infty$) one can pretend that $k_1^\mu$ has been deformed so as to keep momentum conservation valid when $\epsilon k_n^\mu$ vanishes.

The inductive argument assumes that the new system with $n-4$ equations and same number of variables has been solved giving rise to $(n-4)!$ solutions. Let us label the solutions as $\{ \sigma^{(i)}_1,\sigma^{(i)}_2,\ldots ,\sigma^{(i)}_{n-1}\}$ with $i\in \{ 1,2,\ldots ,(n-4)!\}$. Of course, $\sigma^{(i)}_a$ for $a\in \{1,2,3\}$ have values independent of $(i)$ which are fixed from the start. At this point no values for $\sigma_n$ have been found.

In order to proceed, divide the interval $[0,1]$ into $M$ segments. The goal at this point is to find solutions of the system of equations at $\epsilon  = 1/M$. Given that $\epsilon$ is not equal to zero, we can consider the last equation in (\ref{sol}), which after the substitution $s_{nb}\to s_{nb}(\epsilon )$ remains invariant up to an overall factor of $\epsilon$,
\be
\epsilon \sum_{b=2}^{n-1}\frac{s_{nb}}{\sigma_n-\sigma_b}=0. \label{rec}
\ee
It is possible to obtain values of $\sigma_n$ which are very close to the actual solutions at $\epsilon = 1/M$ by simply solving (\ref{rec}) when all variables except $\sigma_n$ are taken to be evaluated on one of the $(n-4)!$ solutions found previously. Extracting the numerator of equation (\ref{rec}) one finds
\be
\sum_{b=2}^{n-1}s_{nb}\prod_{c=2, c\neq b}^{n-1}(\sigma_n-\sigma^{(i)}_c) = 0.
\ee
This is a polynomial of degree $n-3$ in $\sigma_n$ and therefore we get $n-3$ approximate values of $\sigma_n$ for each of the $(n-4)!$ solutions to the remaining $n-1$ $\sigma$'s. Let's denote the approximate values $\sigma_n^{(i,J)}$ with $J=1,\ldots, n-3$.

The next step is to use any standard numerical algorithm\footnote{For example, this is implemented in {\textsc{Mathematica}} as the function {\texttt{FindRoot}}} to find a solution to the full system of equations at $\epsilon = 1/M$ near each of the $(n-3)!$ points $\{ \sigma^{(i)}_1,\sigma^{(i)}_2,\ldots ,\sigma^{(i)}_{n-1}, \sigma_n^{(i,J)} \}$.

Finally, one can iterate the last part of the algorithm solving the equations at $\epsilon = m/M$ using the solutions found at $\epsilon = (m-1)/M$ as seed. In this way one can follow all $(n-3)!$ solutions to the point $\epsilon =1$ and obtain the desired answer. For more details on how this algorithm can be implemented in practice, issues that can arise and some examples including a solution for $n=15$, see appendix \ref{app:numericalalgorithm}.

In three and four dimensions this algorithm can be made more efficient when applied to finding solutions in particular $R$-charge sectors which are not present in higher dimensions but we will not discuss this here.

\section{General Kinematic Invariants and Factorization}~\label{sec:factorization}

In this section we discuss some of the properties of the scattering equations which motivate their physical interpretation as being the backbone of tree level scattering of massless particles.

The most general kinematic invariants in a physical theory are constructed by taking a subset of labels ${\cal S}\subset \{1,2,\ldots ,n\}$ and computing
\be
\left(\sum_{a\in {\cal S}}k^\mu_a\right)^2.
\ee
Without loss of generality one can classify the kinematic invariants by the number of elements in ${\cal S}$ and take a representative in each group given by
\be
(k_1+k_2+\cdots +k_m)^2
\ee
with $m = |{\cal S}|$.

With the benefit of hindsight, let us introduce a special change of variables tailored to the study of this invariant,
\be\label{reparametrization}
\sigma_a=\sigma_n+\tau u_a,\quad\text{with}\quad a\in\{m+1,\ldots,n-1\},
\ee
where we regard $u_{n-1}$ as a fixed value while $\{\tau,u_{m+1},\ldots,u_{n-2}\}$ are new variables.
%

Consider the first $m$ scattering equations
\be
\sum_{b\neq a}\frac{s_{ab}}{\sigma_a-\sigma_b} = 0 \quad {\rm with} \quad a\in \{1,\ldots, m\}.
\ee
At this point it is convenient to introduce a shorthand notation $\sigma_{ab}\equiv \sigma_a-\sigma_b$ (as well as $\sigma_{a,b}$ in case the other notation may cause confusion), which will be used throughout this paper when needed to keep formulas easily readable.

Multiplying the $a^\text{th}$ equation by $\sigma_a-\sigma_n$ and after a simple manipulation we get
\be\label{leftequations}
\frac{\sigma_{a,n}}{\sigma_{a,2}}s_{a,2}+\cdots+\frac{\sigma_{a,n}}{\sigma_{a,m}}s_{a,m}+\left(1+\frac{\tau u_{m+1}}{\sigma_{a,m+1}}\right)s_{a,m+1}+\cdots+\left(1+\frac{\tau u_{n-1}}{\sigma_{a,n-1}}\right)s_{a,n-1}+s_{a,n}=0,
\ee
with $a\in\{1,\ldots,m\}$.

Adding up all these $m$ equations leads to a formula for the kinematic invariant of interest
\be\label{factorizationchannel}
(k_1+\cdots+k_m)^2 =\tau\left(\frac{u_{m+1}}{\sigma_{1,m+1}}s_{1,m+1}+\cdots+\frac{u_{n-1}}{\sigma_{m,n-1}}s_{m,n-1}\right).
\ee

This equation is the core of the connection between the physical space of kinematic invariants and the configuration of puncture locations on $\mathbb{CP}^1$. Note that on the factorization channel where $k_1^\mu+\cdots+k_m^\mu$ is on-shell, i.e.~$(k_1+\cdots+k_m)^2=0$, one has two branches of solutions. One with $\tau=0$ and the other with
\be\label{branch2}
\left(\frac{u_{m+1}}{\sigma_{1,m+1}}s_{1,m+1}+\cdots+\frac{u_{n-1}}{\sigma_{m,n-1}}s_{m,n-1}\right) = 0.
\ee

Consider the solution $\tau=0$. In this branch, the scattering equations separate into two sets of scattering equations but with less particles on each. Very nicely, all the equations in (\ref{leftequations}) directly become
\be\label{leftequations2}
\frac{s_{a2}}{\sigma_{a2}}+\cdots+\frac{s_{am}}{\sigma_{am}}+\frac{s_{aI}}{\sigma_{an}}=0,
\ee
where $k_I^\mu=-(k_1^\mu+\cdots+k_m^\mu)$ is the new null vector.

Any subset of $m-2$ of these equations can be taken as the definition of the scattering equations for the set of $(m+1)$ null vectors $\{ k_1^\mu , k_2^\mu ,\ldots ,k_m^\mu , k_I^\mu \}$ if $\sigma_n$ is identified with $\sigma_I$. In particular, note that by adding all the equations one finds the equation corresponding to $a=I$, i.e.
\be
\sum_{a=1}^m \frac{s_{aI}}{\sigma_{aI}} = 0.
\ee

The second set of scattering equations can simply be obtained by repeating the same analysis after using momentum conservation in the form
\be
(k_1+k_2+\cdots+k_m)^2 = (k_{m+1}+k_{m+2}+\cdots+k_m)^2
\ee
and coordinates $\sigma_a = \sigma_1 + \tau v_a$.

This indicates that the original $(n-3)!$ solutions give rise to $(m-2)!\times (n-m-2)!$ singular solutions in the branch $\tau=0$. Here `singular' refers to the situation when some crossratios of $\sigma_a$'s belong to the set $\{0,1,\infty\}$. It is not difficult to show that crossratios which involve two points from the set $\{1,2,\ldots ,m\}$ and two points from its complement are singular, while all others remain finite.

This is all, of course, very familiar in the study of string scattering amplitudes in which the boundary of the moduli space of Riemann surfaces is connected to the physical singularities of amplitudes \cite{Green:1987sp}. This has also been observed in four dimensions in the context of field theories in twistor space constructions \cite{Vergu:2006np,Skinner:2010cz,Cachazo:2012pz}. These facts are more evidence of the universality of the scattering equations.

\section{KLT Orthogonality}~\label{sec:KLT}

In this section we explore a very striking property of the solutions to the scattering equations which we call KLT orthogonality. In short, the property states that``Parke-Taylor" vectors made from distinct solutions are orthogonal with respect to the Kawai-Lewellen-Tye (KLT) bilinear. The precise meaning of this terminology will be give in subsection \ref{kltorthogonalityproof}. Before moving to the definitions and to the proof of the property let us prove a crucial result which at first sight seems unrelated.

\subsection{Generalized Jacobian and Its Rank}

Let us study the Jacobian matrix associated to the scattering equations. This is a symmetric $n \times n$ matrix with entries,
\be
\Phi_{a b}\equiv \partial\left(\sum_{c\neq a} \frac{s_{a
c}}{\sigma_{a}-\sigma_{c}}\right)/\left(\partial\,\sigma_b\right)=\begin{cases} \displaystyle \frac{s_{a b}}{(\sigma_{a}-\sigma_{b})^2}, \quad a\neq b,\\
\displaystyle -\sum_{c\neq a}\Phi_{a c},\quad a=b.\end{cases}
\label{cases}\ee
As mentioned above only $n-3$ of the scattering equations are linearly independent and therefore the matrix $\Phi$ has rank $n-3$. This matrix was first encountered by Cachazo and Geyer in \cite{Cachazo:2012da} (inspired by a formula for MHV gravity amplitudes found by Hodges in \cite{Hodges:2012ym}) where it played an important role in a formula for gravity amplitudes constructed from gauge theory ones using the KLT construction\footnote{The diagonal terms in the Cachazo-Geyer formula can be trivially simplified to give those in (\ref{cases}) by using the scattering equations.}. This is not accidental as we will see in the next subsection.

Consider now a generalization of $\Phi_{ab}$ which is natural from its origin as the fusion of two gauge theory amplitudes,
\be
\Psi_{ab, a\neq b}\equiv \frac{s_{a b}}{(\sigma_{a}-\sigma_{b})(\sigma'_{a}-\sigma'_{b})}, \quad \Psi_{a a}\equiv
-\sum_{c\neq a} \Psi_{a c}.
\ee
where $\sigma_a$ and $\sigma'_a$ are both assumed to be solutions to the scattering equations. Note that when the two sets $\{\sigma \}$ and $\{\sigma'\}$ are chosen to be the same solution then $\Psi(\{\sigma \},\{\sigma \})=\Phi(\{\sigma \})$. The matrix $\Psi_{ab}$ has also made an appearance very recently in the studies of string amplitudes \cite{Stieberger:2013hza}.

{\bf Proposition 1.} {\it Given two solutions $\{\sigma \}$ and $\{\sigma' \}$ to the scattering equations, the matrix $\Psi(\{\sigma \},\{\sigma' \})$ has rank $n-3$ when $\{\sigma \} = \{\sigma' \}$ and rank $n-4$ when $\{\sigma \} \neq \{\sigma' \}$.}

\vskip0.1in

{\it Proof:} Let us start by taking $\{\sigma \} \neq \{\sigma' \}$. In this case one can construct four null vectors. Explicitly,
 \be
 v_1 = \left(
         \begin{array}{c}
           1 \\
           1 \\
           \vdots \\
           1 \\
         \end{array}
       \right),\quad v_2 = \left(
         \begin{array}{c}
           \sigma_1 \\
           \sigma_2 \\
           \vdots \\
           \sigma_n \\
         \end{array}
       \right),\quad v_3 = \left(
         \begin{array}{c}
           \sigma'_1 \\
           \sigma'_2 \\
           \vdots \\
           \sigma'_n \\
         \end{array}
       \right),\quad v_4 = \left(
         \begin{array}{c}
           \sigma_1\sigma'_1 \\
           \sigma_2\sigma'_2 \\
           \vdots \\
           \sigma_n\sigma'_n \\
         \end{array}
       \right).
 \ee
The fact that $v_1$ is a null vector is trivial as the columns of $\Psi$ add up to zero. Next, we consider $v_2$, the argument is identical for $v_3$. Computing the product $\Psi\, v_2$ one finds
\be \sum^n_{b=1} \Psi_{a b} \sigma_b=-\sum^n_{b=1} \Psi_{a
b} (\sigma_{a}-\sigma_{b})+\sigma_a(\sum^n_{b=1} \Psi_{a b})=-\sum_{b\neq a} \frac{s_{ab}}{\sigma'_{a}-\sigma'_{b}}=0,
\ee
where the last equation holds by the scattering equations. Finally, for $\Psi\, v_4$ one has
\be \sum^n_{b=1} \Psi_{a b} \sigma_b \sigma'_b=\sum^n_{b=1} s_{a b}-\sigma_a \sum_{b\neq a}\frac{s_{ab}}{\sigma'_{a}-\sigma'_{b}}-\sigma'_a \sum_{b\neq a}\frac{s_{ab}}{\sigma_{a}-\sigma_{b}}+\sigma_a\sigma'_a
\sum_{b=1}^n \Psi_{ab}=0,
\ee
where we have used momentum conservation and the scattering equations.

This computation shows that $\Psi(\{\sigma \},\{\sigma' \})$ has four null vectors and therefore rank $n-4$. If the two solutions are taken to be the same then we loose one null vector as $v_2=v_3$. Therefore $\Psi(\{\sigma \},\{\sigma \})$ only has three null vectors given by
\be
 u_1 = \left(
         \begin{array}{c}
           1 \\
           1 \\
           \vdots \\
           1 \\
         \end{array}
       \right),\quad u_2 = \left(
         \begin{array}{c}
           \sigma_1 \\
           \sigma_2 \\
           \vdots \\
           \sigma_n \\
         \end{array}
       \right),\quad  u_3 = \left(
         \begin{array}{c}
           \sigma_1^2 \\
           \sigma_2^2 \\
           \vdots \\
           \sigma_n^2 \\
         \end{array}
       \right),
 \ee
which ends the proof.

\subsection{Proof of KLT Orthogonality}\label{kltorthogonalityproof}

We proceed to the proof of KLT orthogonality. For each solution of the scattering equations one can construct an $n!$ dimensional vector with entries labeled by a permutation $\omega\in S_n$ and defined as
\be
\frac{1}{(\sigma_{\omega(1)}-\sigma_{\omega(2)})(\sigma_{\omega(2)}-\sigma_{\omega(3)})\cdots (\sigma_{\omega(n)}-\sigma_{\omega(1)})}.
\ee
The entries in a given vector are not all independent. In fact, it is obvious that two permutations related by a cyclic transformation have the same entry. Not so obvious is the fact that the set of entries with two labels, say $1$ and $n$, fixed to some positions form a basis in which all other entries can be expressed as linear combinations. The precise linear combinations are known as the Kleiss-Kuijf (KK) relations~\cite{Kleiss:1988ne}. Even less obvious is the fact that using the scattering equations one can further fix the position of a third label, say $n-1$. These relations give rise to the Bern-Carrasco-Johanssen (BCJ) relations of gauge theory amplitudes in four dimensions~\cite{Bern:2008qj}.

The conclusion from this discussion is that all physical information is encoded in $(n-3)!$--dimensional vectors obtained from the $n!$ ones by fixing the position of $1,n-1,n$ and letting $\omega \in S_{n-3}$ permute the rest of the labels. Even after selecting the three labels we still have the choice of where to put them. We will only make use of two choices: $(1,\omega(2), \ldots , \omega(n-2), n-1,n) $ and $(1,\omega(2), \ldots , \omega(n-2), n,n-1)$. Let's denote the corresponding $(n-3)!$--dimensional vectors by
\be
V(\omega ) = \frac{1}{(\sigma_1-\sigma_{\omega(2)})(\sigma_{\omega(2)}-\sigma_{\omega(3)})\cdots (\sigma_{\omega(n-2)}-\sigma_{n-1})(\sigma_{n-1}-\sigma_{n})(\sigma_{n}-\sigma_{1})},
\label{Vr}\ee
and
\be
U(\omega ) = \frac{1}{(\sigma_1-\sigma_{\omega(2)})(\sigma_{\omega(2)}-\sigma_{\omega(3)})\cdots (\sigma_{\omega(n-2)}-\sigma_{n})(\sigma_{n}-\sigma_{n-1})(\sigma_{n-1}-\sigma_{1})}.
\label{Ur}\ee

In this language the Kawai-Lewellen-Tye construction gives rise to a bilinear form
\be  S[\alpha|\beta]=
\prod^{n{-}2}_{i=2}\left(s_{1, \alpha(i)}+\sum^{i{-}1}_{j=2} \theta(\alpha(j), \alpha(i))_{\beta} s_{\alpha(j),\alpha(i)}\right),
\ee
where $\alpha,\beta\in S_{n-3}$, $\theta(i,j)_\beta=1$ if the ordering of $i,j$ is the same in both sequences of labels, $\alpha(2,\ldots,n{-}2)$ and $\beta(2,\ldots,n{-}2)$, and zero otherwise (the convention we use here follows that in~\cite{BjerrumBohr:2010ta}\footnote{The precise notation used in \cite{BjerrumBohr:2010ta} is $S[\alpha|\beta]_1$ reflecting the fact that $1$ was chosen as a pivot.}).

Given any two solutions of the scattering equations,
$$\{\sigma^{(i)}_1,\sigma^{(i)}_2,\ldots ,\sigma^{(i)}_n\} \quad  {\rm and} \quad  \{\sigma^{(j)}_1,\sigma^{(j)}_2,\ldots ,\sigma^{(j)}_n\}$$
(where $(i),(j)$ label the choice of solutions and take values in $\{ 1,2,\ldots ,(n-3)!\}$), define two vectors, $V^{(i)}(\omega)$ and $U^{(j)}(\omega)$, obtained by evaluating (\ref{Vr}) and (\ref{Ur}) on the corresponding solutions. A natural inner product can then be defined as
\be\label{defscalarproduct}
(i,j):= \sum_{\alpha,\beta\in S_{n-3}}V^{(i)}(\alpha)S[\alpha|\beta]U^{(j)}(\beta).
\ee
Now we are ready to formally state KLT Orthogonality in the following proposition.

\vskip0.15in

{\bf Proposition 2 (KLT Orthogonality).} {\it The inner product \eqref{defscalarproduct} satisfies
\be\label{proposition2}
\frac{(i,j)}{(i,i)^{\frac{1}{2}}(j,j)^{\frac{1}{2}}}=\delta_{ij}
\ee
for any values of $i$ and $j$.}

\vskip0.15in

Before proceeding to the proof we first have to connect the KLT bilinear form to the generalized Jacobian defined in the previous subsection (A similar construction has been observed and discussed in~\cite{Stieberger:2013hza}). The starting point is to note that
$$\frac{(i,j)}{(i,i)^{\frac{1}{2}}(j,j)^{\frac{1}{2}}}$$
is clearly invariant under $\SL2C\times \SL2C$ where one group acts on $\{\sigma^{(i)}\}$ while the other acts on $\{\sigma^{(i)}\}$. Let's partially fix both $\SL2C$ redundancies with the convenient choice $\sigma^{(i)}_{n-1}=\sigma^{(j)}_{n}=\infty$ and $\sigma^{(i)}_{n}=\sigma^{(j)}_{n-1}=1$ and define
\be\displaystyle K_n(\{\sigma \},\{\sigma'\})\equiv\!\!\!
\sum_{\alpha,\beta\in S_{n{-}3}}\frac{1}{\sigma_{1,\alpha(2)}\ldots \sigma_{\alpha(n-3),\alpha(n-2)}}S[\alpha|\beta]\frac{1}{\sigma'_{1,\beta(2)}\ldots \sigma'_{\beta(n-3),\beta(n-2)}},\label{bilinear}
\ee
where $\sigma_{ab} = \sigma_a-\sigma_b$. The motivation for this definition is that $K_n$ appears in the numerator of \eqref{proposition2} after the partial gauge fixing, with the identifications $\sigma=\sigma^{(i)}$ and $\sigma'=\sigma^{(j)}$.

It is also convenient to define an auxiliary co-rank one $(n{-}2) \times (n{-}2)$ matrix $\psi^{(n)}$ with entries,
\be \psi_{a b, a\neq b}=\frac{s_{ab}}{\sigma_{ab} \sigma'_{ab}}, \quad \psi_{a a}=-\sum_{b\neq a} \psi_{a b},\ee
for $a,b=1,\ldots,n{-}2$. Since both the rows and columns of $\psi^{(n)}$ add up to zero, it is easy to see that all $(n{-}3) \times (n{-}3)$ minors of $\psi^{(n)}$ are the same, and therefore invariant under any permutations of $1,2,\ldots, n{-}2$. We denote such a minor as $\det'\psi^{(n)}$.



\vskip0.1in

{\bf Proposition 3.} {\it The two functions defined above are identical up to a sign. Explicitly, $K_n(\{\sigma \},\{\sigma'\})=(-1)^n\det'\psi^{(n)}$.}

\vskip0.1in

{\it Proof:} Here we provide a sketch of the proof postponing all details to appendix \ref{app:prop2} for the interested reader\footnote{Stieberger and Taylor have also provided a proof of this connection in \cite{Stieberger:2013hza} as part of the equivalence among several formulas. Our proof differs from theirs in that it is tailored to Proposition 3.}.

%

The main observation is that both $K_n$ and $\det' \psi^{(n)}$ are rational functions of $\sigma_a$'s which can be taken to be completely unconstrained complex variables. In other words, the set $\{ \sigma \}$ is not assumed to be a solution of the scattering equations. For our purposes we choose a given variable, say $\sigma_1$, and define the functions $K_n(\sigma_1)$ and $\det' \psi^{(n)}(\sigma_1)$ by keeping all other variables fixed.

Clearly, both $K_n(\sigma_1)$ and $\det' \psi^{(n)}(\sigma_1)$ vanish as $\sigma_1\to \infty$. Therefore, it is sufficient to show that $K_n(\sigma_1)$ and $(-1)^n\det'\psi^{(n)}(\sigma_1)$ have the same poles and residues at finite values of $\sigma_1$. The only possible pole locations are at $\sigma_1=\sigma_a$ for $a=2,3,\ldots,n{-}2$. Since both functions are manifestly invariant under permutations of $2,3,\ldots,n{-}2$, we only need to consider one of the $n{-}3$ poles, e.g.~that at $\sigma_1=\sigma_2$. In appendix \ref{app:prop2} we show that the residues of $K_n(\sigma_1)$ and of $(-1)^n\det' \psi^{(n)}(\sigma_1)$ at $\sigma_1=\sigma_2$ indeed agree. This concludes the proof that $K_n=(-1)^n\det' \psi^{(n)}$.

\vskip0.1in

{\it Proof of Proposition 2:} Finally we are ready to put all the pieces together and prove KLT orthogonality. With the canonical choice $\sigma^{(i)}_{n-1}=\sigma^{(j)}_{n}=\infty$ and $\sigma^{(i)}_{n}=\sigma^{(j)}_{n-1}=1$, we have
 \be\label{korthogonality}
\frac{(i,j)}{(i,i)^{\frac{1}{2}}(j,j)^{\frac{1}{2}}}=\frac{K_n(\{\sigma^{(i)}\},\{\sigma^{(j)}\})}{K^{\frac{1}{2}}_n(\{\sigma^{(i)}\},\{\sigma^{(i)}\})K^{\frac{1}{2}}_n(\{\sigma^{(j)}\},\{\sigma^{(j)}\})}.\ee
In addition, one finds that the minor of $\psi$ obtained by removing the first row and column is identical to that of $\Psi(\{\sigma \},\{\sigma' \})$ after removing rows $\{1,n-1,n\}$ and columns $\{1,n-1,n\}$ in this canonical gauge. We denote these determinants by $|\psi^{(n)}|^1_1$ and $|\Psi|_{1,n-1,n}^{1,n-1,n}$ respectively. Using first Proposition 3 the following identity holds,
\be
\frac{(i,j)}{(i,i)^{\frac{1}{2}}(j,j)^{\frac{1}{2}}}=\frac{|\Psi(\{\sigma^{(i)}\},\{\sigma^{(j)}\})|_{1,n-1,n}^{1,n-1,n}}{(|\Psi(\{\sigma^{(i)},\sigma^{(i)}\}|_{1,n-1,n}^{1,n-1,n})^{\frac 12}|(\Psi(\{\sigma^{(j)}\},\{\sigma^{(j)}\})|_{1,n-1,n}^{1,n-1,n})^{\frac 12}}~\label{KLTorthogonal}
\ee
and using Proposition 1, we learn that the denominators are not singular since $\Psi(\{\sigma^{(k)}\},\{\sigma^{(k)}\})$ has co-rank $3$ while the numerator vanishes if $i\neq j$ since $\Psi(\{\sigma^{(i)}\},\{\sigma^{(j)}\})$ has co-rank four.
This concludes our proof of KLT orthogonality.

\section{Discussions}~\label{sec:discussion}

We end with some comments on possible future directions related to applications of the scattering equations and KLT orthogonality to tree--level amplitudes of massless particles in arbitrary dimensions as well as a very intriguing connection to the high energy scattering of strings.

\subsection{Towards Yang-Mills and Gravity Amplitudes in Arbitrary Dimensions}

In the light of the properties discussed in this paper it is very tempting to propose the existence of formulas for Yang-Mills and gravity scattering amplitudes in any dimensions based on rational maps. The first important observation is the construction of an $\SL2C$ covariant and permutation invariant measure that restricts an integration over $\sigma_a$ to solutions to the scattering equations. This is easily achieved by noting that
\be
\prod_a {}'\delta(\sum_{b\neq a} \frac{s_{a b}}{\sigma_{a b}}) \equiv \sigma_{ij}\sigma_{jk}\sigma_{ki}\prod_{a\neq i,j,k}\delta(\sum_{b\neq a} \frac{s_{a b}}{\sigma_{a b}})
\ee
is independent of the choice $\{i,j,k\}$ and hence permutation invariant (Recall that $\sigma_{ab} = \sigma_a-\sigma_b$).

With this observation, it is natural to propose the following formulations for gauge theory and gravity amplitudes in any dimensions
\ba M^{\mathrm{YM}}_n(1,2,\ldots,n)&=&\int \frac{d\,^n\sigma}{\textrm{vol}\,\SL2C} \prod_a {}'\delta(\sum_{b\neq a} \frac{s_{a b}}{\sigma_{a b}}) \frac{E(\{k,\epsilon,
\sigma\})}{\sigma_{12}\ldots\sigma_{n1}},~\label{YM}\\ M^{\mathrm{gravity}}_n&=&\int\frac{d\,^n\sigma}{\textrm{vol}\,\SL2C} \prod_a {}'\delta(\sum_{b\neq a} \frac{s_{a
b}}{\sigma_{a b}}) E^2(\{k,\epsilon, \sigma\}), ~\label{gravity}
\ea
where $E(\{k,\epsilon, \sigma\})$ is a permutation invariant function of $\sigma_a$, momenta $k_a^\mu$ and polarization vectors $\epsilon_a^\mu$.
Note that $\SL2C$ invariance of the integrand constraints $E$: under an $\SL2C$ transformation, $\sigma_a\to \frac{A\sigma_a+B}{C\sigma_a+D}$, $E$ must transform as
\be
E(\{k,\epsilon,
\sigma\})\to E(\{k,\epsilon,
\sigma\}) \prod_{a=1}^n(C\sigma_a+D)^2.
\ee
Most likely, $E$ should also be gauge invariant for each solution to the scattering equations. This possibility motivates the idea of considering the contribution from each solution as building blocks. Each block satisfies many physical properties that the full amplitude satisfy. We leave the investigation of this fascinating possibility for future work.

Some evidence suggesting that the formula for Yang-Mills must exist is the following. Firstly, similar formulas in four dimensions can be derived for individual R-charge sectors. Secondly, the BCJ fundamental relations for Yang-Mills amplitudes are known to hold in any dimension~\cite{Bern:2008qj,BjerrumBohr:2009rd}. Using the arguments in \cite{Cachazo:2012uq} one can show that $M^{\mathrm{YM}}_n$ satisfies them as a direct consequence of the scattering equations.

It is clear from KLT orthogonality and the KLT relations that, if the formula for Yang-Mills exists, so does the formula for gravity. In a nutshell, define the permutation invariant combination $\det'\Psi(\{\sigma\},\{\sigma'\})\equiv |\Psi(\{\sigma\},\{\sigma'\})|^{i j k}_{r s t}/(\sigma_{i j}\sigma_{j k}\sigma_{k i} \sigma_{r s}\sigma_{s t}\sigma_{t r})$, and note that when $\{\sigma\}=\{\sigma'\}$ it reduces to the permutation invariant Jacobian of scattering equations $\det'\Phi(\{\sigma\})\equiv \det'\Psi(\{\sigma\},\{\sigma\})$. The KLT relations for two copies of Yang-Mills amplitudes give,
\ba
&& \!\!\!\!\int\frac{d\,^n \sigma}{\textrm{vol}\,\SL2C} \prod_a {}' \delta(\sum_{b\neq a} \frac{s_{a b}}{\sigma_{a b}})\int\frac{d\,^n \sigma'}{\textrm{vol}\,\SL2C} \prod_a {}' \delta(\sum_{b\neq a} \frac{s_{a b}}{\sigma'_{a b}})\det{}'\Psi(\{\sigma\}, \{\sigma'\})E(\{\sigma\})E(\{\sigma'\})\nl
=&& \!\!\! \sum^{(n{-}3)!}_{i,j=1}\frac{\det{}'\Psi(\{\sigma^{(i)}\},\{\sigma^{(j)}\})} {\det{}'\Phi(\{\sigma^{(i)}\})\det{}'\Phi(\{\sigma^{(j)}\})}E(\{\sigma^{(i)}\})E(\{\sigma^{(j)}\})\nl
=&& \int\frac{d\,^n \sigma}{\textrm{vol}\,\SL2C} \prod_a {}' \delta(\sum_{b\neq a} \frac{s_{a b}}{\sigma_{a b}}) E^2(\{\sigma\}),
\ea
where we have denoted the $i$-th solution as $\sigma^{(i)}$ for $i\in \{1,\ldots,(n{-}3)!\}$, and the last equality follows from Proposition 2.

\subsection{High Energy Scattering of Strings}~\label{sec:string}

Recently~\cite{Mafra:2011nv,Mafra:2011nw} it was shown that superstring disk amplitudes can be expressed as linear combinations of $(n{-}3)!$ Yang-Mills partial amplitudes (e.g. in the $U$ basis defined in \eqref{Ur}) with coefficients encoding $\alpha'$-corrections,
\ba M^{\mathrm{open}}_n (\alpha')&=&\sum_{\tau,\rho\in S_{n{-}3}}\int\frac{d^n z_i}{\textrm{vol\,PSL}(2,\mathbb{R})} \prod_{i< j} |z_{i j}|^{\alpha' s_{i j}} \frac{S[\rho|\tau]}{z_{1,\rho(2)} \ldots z_{n{-}1,n} z_{n,1}}M_n^{\mathrm{YM}}(\tau)\nl
&\equiv& \int D^{n{-}3}_{\alpha'} z \sum_{\rho\in S_{n{-}3}}\frac{S[\rho|\tau]}{z_{1,\rho(2)} \ldots z_{n{-}1,n} z_{n,1}}M^{\mathrm{YM}}_n(1,\tau(2,\cdots,n{-}2), n, n{-}1),~\label{open1}\ea
where one integrates over positions of string vertex operators on the disk boundary, with the canonical ordering $z_i<z_{i{+}1}$, and fixes the PSL$(2,\mathbb{R})$ redundancy by choosing e.g. $(z_1,z_{n{-}1}, z_n)=(0,1,\infty)$; $z_{ij}\equiv z_i-z_j$ and the ``Parke-Taylor" vectors of $z$'s are in the $V$ basis. On the first line we have denoted $M^{\mathrm{YM}} (\tau)\equiv M^{\mathrm{YM}}(1,\tau(2,\cdots,n{-}2),n,n{-}1)$); on the second line, we have denoted the disk integral measure with Koba-Nileson factors as $D^{n{-}3}_{\alpha'} z$, which contains the entire $\alpha'$-dependence.

Closed string tree amplitudes are given by integrals over string vertex insertions $ (z_i, \bar z_i) $ on a complex sphere. It is well known that by decomposing vertex operators
the integrand can be written as the product of left-moving and right-moving open string integrands~\cite{Kawai:1985xq}, thus the closed string amplitudes are bilinear forms of two copies of Yang-Mills amplitudes,
\be
M^{\mathrm{closed}}_n (\alpha')=\! \int D^{n{-}3}_{\alpha'}z_i D^{n{-}3}_{\alpha'}\bar z_i \!\!\!\sum_{\tau,\tilde\tau, \rho,\tilde\rho} \frac{S[\rho|\tau]S[\tilde\rho|\tilde\tau]}{z_{1,\rho(2)} \ldots z_{n{-}1,n} z_{n,1}\bar z_{1,\tilde\rho(2)} \ldots \bar z_{n{-}1,n} \bar z_{n,1}} M^{\mathrm{YM}}_n (\tau) M^{\mathrm{YM}}_n(\tilde\tau) ~\label{closed1}\ee
where the sum is over $\tau,\tilde\tau, \rho,\tilde\rho\in S_{n{-}3}$.

Plugging \eqref{YM} into \eqref{open1} and \eqref{closed1}, and using Proposition 2 we find,
\ba &&M^{\mathrm{open}}_n (1,\ldots, n;\alpha')= \int_{z_i<z_{i{+}1} }D^{n{-}3}_{\alpha'} z_i \int  \frac{d^n \sigma}{\textrm{vol}\,\SL2C}\prod_a{}' \delta(\sum_{b\neq a}\frac{s_{ab}}{\sigma_{ab}})\,E(\{\sigma\}) 
\det{}' \Psi(\{z\}, \{\sigma\}),\nl
&& M^{\mathrm{closed}}_n (\alpha')=\int_{\mathbb{C}^{n{-}3}}  D^{n{-}3}_{\alpha'} z_i   D^{n{-}3}_{\alpha'} \bar z_i \int  \frac{d^n \sigma}{\textrm{vol}\,\SL2C}\prod_a{}' \delta(\sum_{b\neq a}\frac{s_{ab}}{\sigma_{ab}}) \frac{d^n \tilde \sigma}{\textrm{vol}\,\SL2C}\prod_a{}' \delta(\sum_{b\neq a}\frac{s_{ab}}{\tilde\sigma_{ab}})\nl
&& \times E(\{\sigma\})E(\{\tilde\sigma\})\det{}' \Psi(\{z\},\{\sigma\})\det{}' \Psi(\{\bar z\}, \{\tilde \sigma\}).\label{openclosed}
\ea

In \eqref{openclosed}, the open string tree amplitude is quite nicely given by a sum of disk-boundary integrals of $\det'\Psi$, over $(n{-}3)!$ solutions; in contrast, for closed string amplitudes, in addition to sphere integrals of two copies of $\det'\Psi$, one needs to sum over $(n{-}3)!^2$ solutions! This can be compared with gravity amplitudes, \eqref{gravity}, where by KLT orthogonality one sums over only $(n{-}3)!$ terms. In particular, in maximal supergravity in four dimensions, tree amplitudes enjoy an enhanced R-symmetry $SU(8)$ due to KLT orthogonality, but it is well known that closed-string amplitudes only have $SU(4)\times SU(4)$ symmetry.

In~\cite{Gross:1987ar} the high energy limit, \emph{i.e.} the limit $M^2\gg 1/\alpha'$ where kinematic invariants have a typical energy scale, $s_{ab}\sim M^2$, was considered, and disk/sphere integrals can be evaluated using saddle point approximations. We observe that the saddle point equations are exactly the scattering equations we found in field theory. For disk integrals, $z$'s are real, and we have
\be
\prod_{a<b}|z_{ab}|^{\alpha's_{a b}}=e^{\frac 1 2\sum_{a\neq b} \alpha' s_{a b}\ln |z_{ab}|}\,\xrightarrow{\smash{\textrm{saddle point}}} \,0=\frac{\partial}{\partial z_a} \sum_{b\neq a}s_{ab}\ln|z_{ab}|=\sum_{b\neq a} \frac{s_{ab}}{z_{ab}},
\ee
for $a=1,\ldots, n$. For sphere integrals, we have $\ln |z_{ab}|^2=\ln (z_{ab}\bar z_{ab})$ instead of $\ln |z_{ab}|$. Note that all the kinematic invariants are real, the saddle point equations for $z$ and those for $\bar z$ are equivalent. In the limit $\alpha' s_{ab}\sim \alpha' M^2 \to \infty$, the sphere integral can be approximated by a sum over saddle points labeled by $i\in\{1,2,\ldots,(n{-}3)!\}$. Explicitly, we have that $ M^{\mathrm{closed}}_n$ can be written as
\ba & & \sum_{i,j,\tilde j=1}^{(n{-}3)!} \frac{\prod_{a<b}|z_{ab}^{(i)}|^{ \alpha' s_{ab}}}{{\det{}'\Phi(z^{(i)})}^{\frac{1}{2}}{\det{}'\Phi(\bar{z}^{(i)})}^{\frac{1}{2}}}
\frac{\det{}'\Psi(z^{(i)},\sigma^{(j)}) \det{}'\Psi(\bar{z}^{(i)},\tilde\sigma^{(\tilde j)})}{\det{}'\Phi(\sigma^{(j)})\det{}'\Phi(\tilde\sigma^{(\tilde j)})}
E(\sigma^{(j)})E(\tilde{\sigma}^{(\tilde j)})
+\mathcal{O}(\frac 1{M^2 \alpha'})\nl
& = & \sum_{i=1}^{(n{-}3)!} \frac{\prod_{a<b} |\sigma_{ab}^{(i)}|^{\alpha' s_{ab}}}
{{\det{}'\Phi(\sigma^{(i)})}^{\frac{1}{2}}{\det{}'\Phi(\bar{\sigma}^{(i)})}^{\frac{1}{2}}} E^2(\{k,\varepsilon,\sigma^{(i)}\})+\mathcal{O}(\frac 1{M^2 \alpha'}),
\ea
where on the second line we identified $\sigma^{(i)}_a=z^{(i)}_a$, $\tilde\sigma^{(i)}_a=\bar z^{(i)}_a$ using Proposition 2, and the leading term becomes a sum over $(n{-}3)!$ solutions. Hence we see that simplifications of closed string amplitudes in the high energy limit are closely related to the KLT orthogonality \eqref{proposition2}. It is very intriguing that the high energy limit related to $\alpha'\to \infty$ bears significant similarities with the field theory limit, $\alpha'\to 0$. It would be fascinating to further explore potential connections.

\acknowledgments

The authors would like to thank Erik Schnetter for useful discussions on the numerical algorithms. This work is supported by the Perimeter Institute for Theoretical Physics. Research at Perimeter Institute is supported by the Government of Canada through Industry Canada and by the Province of Ontario through the Ministry of Research \& Innovation.

\emph{Note}: While this manuscript was being prepared for submission, the paper~\cite{Stieberger:2013nha} appeared which has overlap with some of the results of this work.

\appendix

\section{Numerical Algorithm}\label{app:numericalalgorithm}

In this appendix we discuss some details of the numerical implementation of the algorithm described in section 3 for finding solutions of the scattering equations.

In physical applications one is usually interested in real kinematic data, i.e., $s_{ab}\in \mathbb{R}$. In this case, it is likely that when taking values of $\epsilon$ in the interval $[0,1]$ one can come close to a singularity of the Jacobian of the equations. This is undesirable as most numerical algorithms use Newton's method and could fail. In this case a more convenient choice is to use a unit semicircle centered at $\epsilon = 1/2$ in the complex $\epsilon$ plane. This technique also has the advantage of avoiding falling into a cycle outside the basin of attraction of the solution.

In theoretical computations one usually chooses rational kinematic data, i.e., $s_{ab}\in \mathbb{Q}$. The reason is that physical amplitudes are computed as a sum of a rational function of $\sigma$'s evaluated over all $(n-3)!$ solutions and therefore it is guaranteed to be a rational number. Using enough working precision one can rationalize and hence obtain the exact answer even though the intermediate steps were numerical. In this particular case, an additional issue can arise; using a large number of rational kinematic invariants increases the possibilities of simple linear relations arising among them and therefore the number of possible Jacobian singularities. In this case one should start with rational numbers that are unlikely to satisfy simple linear relations. For example, instead of generating rational numbers for $s_{ab}$ by using the function {\texttt{RandomInteger}}$[\{-100,100\}]/100$ in {\textsc{Mathematica}} one should use {\texttt{RandomInteger}}$[\{-10^6,10^6\}]/10^6$.

\subsection{A More Explicit Version of the Algorithm}

In order to implement algorithm in practice one can start by defining the equations as follows
\be
\sum_{b\neq a}\frac{\epsilon_a\epsilon_b s_{ab}}{\sigma_a-\sigma_b}=0,\quad a\in\{4,5,\ldots,n\}.
\ee
A good starting point for the algorithm is to deform the system of equations for $n$ particles down to seven points where all solutions can be easily found using, e.g., {\texttt{NSolve}} in \textsc{Mathematica}. This is achieved by simply taking
\be
\epsilon_a = \begin{cases} \displaystyle 1 \quad a \in \{ 1,2,\ldots, 7\}\\
\displaystyle 0\quad a\in \{ 8,9,\ldots, n \},\end{cases}
\ee
This system has $24$ solutions and the algorithm can then be applied by letting $\epsilon_8=1/10^3$ and proceeding as in section 3. This will lead to $5$ solutions for each of the original $24$ solutions at $\epsilon_8 =1$. This can then be continued until reaching the desired $(n-3)!$ solutions. It is important to mention that in this implementation, one does not have to perform each step to a very high working precision. The reason is that once $(n-3)!$ approximate solutions have been found, one can use them as seeds in a {\texttt{FindRoot}} procedure with the desired working precision. This is actually welcome as the function {\texttt{FindRoot}} can fail to reach the default precision goals for intermediate values of $\epsilon$ but this is not important as long as one can continue to get an approximate value of a solution at $\epsilon =1$.

\subsection{Example}

Here we give as an example the result of the computation of one solution for $n=15$. The kinematic data was chosen as random rational numbers for $s_{ab}$ with $1 \leq a < b \leq 14$ given below in lexicographic order
$$
\left\{\frac{49}{13},-\frac{69}{26},-\frac{17}{6},-\frac{29}{25},
\frac{52}{11},\frac{55}{3},-\frac{43}{34},-5,-\frac{5}{16},-\frac{76}{45},-\frac{8}{43},
\frac{73}{45},-\frac{15}{47},-\frac{25}{3},\frac{90}{37},2,-24,\frac{35}{24},56,\right.
$$

$$ -\frac{49}{23},
-\frac{20}{3},\frac{1}{11},\frac{16}{41},\frac{31}{16},-\frac{31}{7},\frac{79}{19},
-\frac{49}{25},-\frac{58}{7},-\frac{10}{3},\frac{16}{7},\frac{9}{11},\frac{21}{10},\frac{71}{21},
\frac{17}{19},-\frac{47}{12},\frac{5}{24},-\frac{31}{48},-\frac{65}{19},
$$

$$ \frac{32}{37},-\frac{85}{41},
-\frac{35}{13},-\frac{79}{44},\frac{7}{5},-\frac{9}{5},\frac{19}{5},\frac{23}{15},-\frac{1}{2},\frac{74}{11},
-\frac{38}{43},-\frac{56}{9},-\frac{13}{4},-\frac{7}{4},-\frac{39}{35},\frac{11}{7},-\frac{3}{34},\frac{21}{22},
\frac{56}{23},$$

$$\frac{78}{35},\frac{22}{7},\frac{6}{13},-\frac{48}{25},-\frac{49}{43},\frac{4}{33},-\frac{5}{4},
-\frac{39}{23},\frac{29}{28},\frac{16}{7},-\frac{15}{8},\frac{25}{12},-\frac{21}{25},-\frac{95}{14},
-\frac{7}{5},\frac{5}{8},\frac{17}{36},\frac{31}{43},7,\frac{42}{31},\frac{33}{43},
$$

$$\left. \frac{80}{23},
-\frac{39}{38},\frac{27}{17},\frac{8}{7},-\frac{11}{27},\frac{94}{27},-\frac{11}{8},9,-\frac{43}{20},
\frac{82}{33},-\frac{61}{6},\frac{10}{33} \right\} \qquad\qquad\qquad\qquad\qquad\qquad\qquad\qquad
$$
and with
\be
s_{13,14} = -\sum_{1<a<b<14}s_{ab}
\ee
The rest of the kinematic invariants, i.e., those of the form $s_{15,a}$, are obtained by using momentum conservation. The data does not satisfy any Gram determinant equations and therefore it can be taken to correspond to general dimensions. Note that working in general dimensions means that this computation is at least equivalent to that of a computation in a four dimensional $k=7$ or N${}^5$MHV sector.

Below we show the solution to $25$ digits of precision but using it as a seed one can increase the precision at very little cost.

Recall that using the $\SL2C$ invariance of the equations we choose to fix the value of the first three $\sigma$'s. Our choice is $\sigma_1\to \infty$, $\sigma_2 =0$ and $\sigma_3=1$. The remaining $12$ $\sigma$'s are:
$$
\begin{array}{ccl}
\sigma_4 & = & 0.5496193504318654421011817+0.3319870690807634365609033 i \\
\sigma_5 & = & 0.13759872147189287197217158+0.00723562043413197816599421 i \\
\sigma_6 & = &  0.6947313566266571926077990+0.0526139619994048265322461 i \\
\sigma_7 & = & 1.0865440087837388144794260+0.3768740374717843745501630 i \\
\sigma_8 & = &  23.33664228765061141631063-0.33658205624253639217240 i \\
\sigma_9 & = &  2.160942080177577331147289+1.124899577655694226591314 i \\
\sigma_{10} & = &  3.265453266150672900415255-3.358839499723287173276593 i \\
\sigma_{11}  & = & 0.04841609220705573133055635+0.00901941920135424505351115 i \\
\sigma_{12}  & = & 3.124269149584940975154808-2.943058583858387329266730 i \\
\sigma_{13} & = & 1.1676848034334310319712031-0.0732151746711586090152780 i \\
\sigma_{14} & = &  -19.36286341034387769858701+4.73029434693411454422806 i \\
\sigma_{15} & = &  -1.1581504995107345733001948+0.0739431466630364554298497 i. \\
\end{array}
$$

\section{Details of the Proof of Proposition 3}~\label{app:prop2}

In the proof of KLT orthogonality (Proposition 2), one important ingredient is Proposition 3, which states that
\be\label{app:prop3}
K_n=(-1)^n{\det}'\psi^{(i)},
\ee
where
\be
K_n:=\sum_{\alpha,\beta\in S_{n-3}}\frac{1}{\sigma_{1,\alpha(2)}\cdots\sigma_{\alpha(n-3),\alpha(n-2)}}
S[\alpha|\beta]\frac{1}{\sigma'_{1,\beta(2)}\cdots\sigma'_{\beta(n-3),\beta(n-2)}},
\label{ares}\ee
and $\psi^{(n)}$ is a co-rank one $(n-2)\times(n-2)$ matrix
\be
\psi_{ab}:=\begin{cases}\frac{s_{ab}}{\sigma_{ab}\sigma'_{ab}},&a\neq b\\-\sum_{c\neq a}\psi_{ac},&a=b\end{cases},
\ee
where $a,b\in \{1,2,\ldots ,n-2\}$. And $\det'$ is any $(n-3)\times(n-3)$ minor of $\psi^{(n)}$. Here both permutations $\alpha$ and $\beta$ act on the set of labels $\{2,3,\ldots,n-2\}$. In the formula above we use the natural convention in which the label $i$ in $\beta(i)$ denotes exactly the position in the whole sequence $(1,\beta(2),\ldots,\beta(n-2))$. The reason we mention this is that below the proof requires the introduction of other permutations with different properties.

The first step in the proof of \eqref{app:prop3} is to study all terms in \eqref{ares} with a given permutation $\alpha$. This motivates the introduction of
\be
X'_n(1,\alpha(2),\ldots,\alpha(n-2)):=\sum_{\beta\in S_{n-3}}\frac{S[\alpha|\beta]}{\sigma'_{1,\beta(2)}\cdots\sigma'_{\beta(n-3),\beta(n-2)}}
\label{jis}\ee
so that \eqref{ares} becomes
\be\label{app:KnXn}
K_n=\sum_{\alpha\in S_{n{-}3}}\frac{X'_n(1,\alpha(2),\ldots,\alpha(n{-}2))}{\sigma_{1,\alpha (2)}\ldots \sigma_{\alpha(n-3),\alpha(n-2)}}.
\ee

\subsection{Preliminary Simplification of $K_n$}

Next we prove the identity
\be\label{app:Xn}
X'_n(1,\alpha(2),\ldots,\alpha(n-2))=
\prod_{i=2}^{n-2}\sum_{j=1}^{i-1}\frac{s_{\alpha(j),\alpha(i)}}{\sigma'_{\alpha(j),\alpha(i)}}.
\ee
This is done recursively. Without loss of generality, we can fix $\alpha$ to be the identity permutation, i.e., $\alpha(i)=i$.

Firstly \eqref{app:Xn} is trivially correct at $n=4$. When $n>4$, we assume that \eqref{app:Xn} holds for the case $n-1$, and show that $X'_n(1,2,\ldots,n-2)$ defined by \eqref{jis} satisfies the recursion relation
\be\label{app:Xnrecursion}
X'_n(1,2,\ldots,n-2)=X'_{n-1}(1,2,\ldots,n-3)\sum_{m=1}^{n-3}\frac{s_{m,n-2}}{\sigma'_{m,n-2}},
\ee
which obviously holds for the r.h.s.~of \eqref{app:Xn}.

Clearly $\sigma'_{n-2}$ only appears in the factor multiplying $X'_{n-1}$ on the r.h.s.~of \eqref{app:Xnrecursion}, so the strategy for proving the recursion is simply to collect terms in $X'_{n}$ according to factors of the form $1/\sigma'_{m,n-2}$ for every $m\in\{1,\ldots,n-3\}$. In this operation label $n-2$ is special, therefore it is convenient to define permutations $\gamma\in S_{n-4}$ which act on the reduced label set $\{2,\ldots,n-3\}$. Moreover, we are interested in collecting terms where $n-2$ sits at a particular position, say $\beta(k) =n-2$. For terms of this form we define
\be
\gamma(i)=\begin{cases}\beta(i),&i<k\\\beta(i+1),&i\geq k.\end{cases}
\ee
Next, modify the terms in $X'_n$ for which $k<n-2$ (i.e.~the label $n-2$ does not sit at the end of the original sequence) as follows
\be
\frac{1}{\cdots\sigma'_{\gamma(k-1),n-2}\sigma'_{n-2,\gamma(k)}\cdots}=\frac{1}{\cdots\sigma'_{\gamma(k-1),\gamma(k)}\cdots}\left(\frac{1}{\sigma'_{\gamma(k-1),n-2}}-\frac{1}{\sigma'_{\gamma(k),n-2}}\right),
\ee
where the common factor on r.h.s.~above is exactly the one naturally associated to the reduced sequence $(1,\gamma(2),\ldots,\gamma(n-3))$. As a result of this manipulation, every term in $X'_n$ possesses a unique factor $1/\sigma'_{m,n-2}$ for some $m\in\{1,\ldots,n-3\}$.

We first collect terms with $1/\sigma'_{1,n-2}$, which can be observed to always have the form
\be
\frac{S[2,\ldots,n-2|n-2,\gamma(2),\ldots,\gamma(n-3)]}{\sigma'_{1,\gamma(2)}\cdots\sigma'_{\gamma(n-4),\gamma(n-3)}}\frac{1}{\sigma'_{1,n-2}}.
\ee
Since
\be
S[2,\ldots,n-2|n-2,\gamma(2),\ldots,\gamma(n-3)]=s_{1,n-2}S[2,\ldots,n-3|\gamma(2),\ldots,\gamma(n-3)],
\ee
they directly add up to be
\be\label{app:collection1}
X'_{n-1}(1,\ldots,n-3)\frac{s_{1,n-2}}{\sigma'_{1,n-2}}.
\ee
Then we collect terms with $1/\sigma'_{m,n-2}$ for any $m\in\{2,\ldots,n-3\}$. Observe that for a particular sequence $(1,\gamma(2),\ldots,\gamma(n-3))$, together with a particular identification $m=\gamma(l)$ for some $l\in\{2,\ldots,n-3\}$, the contributing terms always come in pairs, with a common denominator $\sigma'_{1,\gamma(2)}\cdots\sigma'_{\gamma(n-4),\gamma(n-3)}\sigma'_{m,n-2}$, and it is easy to see that the two numerators add up to be
\be
\begin{split}
S[2,\ldots,n-2|\ldots,\gamma(l),n-2,\ldots]-S[2,&\ldots,n-2|\ldots,n-2,\gamma(l),\ldots]\\
=&s_{m,n-2}S[2,\dots,n-3|\ldots,\gamma(l),\ldots].
\end{split}
\ee
Note that the permutation $\gamma$ in the final object $S[2,\ldots,n-3|\gamma]$ exactly matches that associated with the denominator. Considering all possible $\gamma\in S_{n-4}$, it is straightforward to see that all terms which contain the factor $1/\sigma'_{m,n-2}$ add up to
\be\label{app:collection2}
X'_{n-1}(1,\ldots,n-3)\frac{s_{m,n-2}}{\sigma'_{m,n-2}}.
\ee
Combining both \eqref{app:collection1} and \eqref{app:collection2}, we can directly see that the recursion \eqref{app:Xnrecursion} holds, thus verifying \eqref{app:Xn}.

\subsection{Residue Computation}

In order to complete the proof of Proposition 3, simply note that both $K_n$ and $\det' \psi^{(n)}$ are rational functions of $\sigma_1$. In this part of the proof the set $\{ \sigma \}$ is not assumed to be a solution of the scattering equations and therefore $\sigma_1$ is a completely unconstrained complex variable.

A simple observation is that both $K_n(\sigma_1)$ and $\det' \psi^{(n)}(\sigma_1)$ vanish as $\sigma_1\to \infty$. Therefore, it is sufficient to show that $K_n(\sigma_1)$ and $(-1)^n\det'\psi^{(n)}(\sigma_1)$ have the same poles and residues at finite values of $\sigma_1$. Clearly, the only possible pole locations are at $\sigma_1=\sigma_a$ for $a=2,3,\ldots,n{-}2$. Since both functions are manifestly invariant under permutations of $2,3,\ldots,n{-}2$, we only need to consider one of the $n{-}3$ poles, e.g.~that at $\sigma_1=\sigma_2$.

Let us explicitly compute the residue at $\sigma_1=\sigma_2$ for both $K_n(\sigma_1)$ and of $(-1)^n\det' \psi^{(n)}(\sigma_1)$.

From \eqref{app:KnXn} together with \eqref{app:Xn}, it is easy to compute the residue of $K_n(\sigma_1)$ at $\sigma_1=\sigma_2$, which gives,
\be \textrm{Res}_{\sigma_1=\sigma_2} K_n(\sigma_1)=\frac{s_{12}}{\sigma'_{12}}\sum_{\alpha\in S_{n{-}4}}\prod_{a=3}^{n{-}2}\frac 1{\sigma_{\alpha (a{-}1),\alpha (a)}}\sum^{a{-}1}_{b=1}\frac{s_{\alpha(b),\alpha(a)}}{\sigma'_{\alpha(b),\alpha(a)}},\ee
thus we only need to prove that the residue of $\det'\psi^{(n)}(\sigma_1)$ at $\sigma_1=\sigma_2$ is the same.

Given that proving the equality between the residues amounts to proving the equality of the two functions, here we proceed by induction in the statement $K_n(\sigma_1)= (-1)^n\det' \psi^{(n)}(\sigma_1)$. In other words, we check that it is valid for $n=4$ and assume its validity for $n-1$ to show that for the case $n$ both functions have the same residue at $\sigma_1 = \sigma_2$.

The starting point, $K_4=\det'\psi^{(4)}=\frac {s_{12}}{\sigma_{12}\sigma'_{12}}$ is easily checked by inspection.

For $\det'\psi^{(n)}$, we choose to delete the first row and column and denote it as $|\psi^{(n)}|^1_1$. In the minor the only term that has the pole $\sigma_{1}=\sigma_2$ is the entry $\psi_{2,2}$, and we find
\be
\textrm{Res}_{\sigma_1=\sigma_2}\det{}'\psi^{(n)}(\sigma_1)=-\frac{s_{12}}{\sigma'_{12}}|\psi^{(n)}(\sigma_1=\sigma_2)|^{12}_{12},
\ee
where $|\psi^{(n)}|^{12}_{12}$ denotes the minor obtained by deleting rows and columns $1$ and $2$.

Note that when evaluated at $\sigma_1=\sigma_2$,
$$ \psi_{1a}+\psi_{2a}=\frac1 {\sigma_{1a}}(\frac{s_{1a}}{\sigma'_{1a}}+\frac{s_{2a}}{\sigma'_{2a}}),$$
and the minor $|\psi^{(n)}(\sigma_1=\sigma_2)|^{12}_{12}$ can be rewritten as a minor of the following $n{-}3$ by $n{-}3$ matrix $\hat\psi^{(n{-}1)}$,
\be
\hat\psi_{1 c}=\hat\psi_{c 1}=\frac{1}{\sigma_{1c}}(\frac{s_{1 c}}{\sigma'_{1 c}}+\frac{s_{2 c}}{\sigma'_{2 c}}),\quad\hat\psi_{ab,a\neq b}=\psi_{ab},\quad \hat\psi_{aa}=-\sum_{b\neq a,b\neq 2}\hat\psi_{ab},
\ee
for $c=3,4,\ldots,n$ and $a,b=1,3,4,\ldots,n{-}2$, thus $|\psi^{(n)}(\sigma_1=\sigma_2)|^{12}_{12}=|\hat\psi^{(n{-}1)}|^1_1$. Since $|\hat\psi^{(n{-}1)}|^1_1$ is invariant under the permutations of the labels $1,3,\ldots,n{-}2$, by the induction assumption, it is the same function as $(-1)^{n-1}\hat K_{n{-}1}$ given by \eqref{app:KnXn}, in which $\hat X'_{n{-}1}$ is given by \eqref{app:Xn},
\be \hat X'_{n{-}1}(1,3,\ldots,{n{-}2})=\prod_{i=3}^{n{-}2}(\frac{\hat{s}_{1i}}{\sigma'_{1i}}+\sum_{j=3}^{i{-}1}\frac{s_{ji}}{\sigma'_{ji}})=\prod_{i=3}^{n{-}2}\sum_{j=1}^{i{-}1}\frac{s_{ji}}{\sigma'_{ji}}\ee
and its permutations on the set of $3,\ldots,n{-}2$. Thus we conclude that
\be
\textrm{Res}_{\sigma_1=\sigma_2}\det{}'\psi^{(n)}(\sigma_1)=(-1)^{n}\frac{s_{12}}{\sigma'_{12}}\sum_{\alpha\in S_{n{-}4}}\prod_{i=3}^{n{-}2}\frac1 {\sigma_{\alpha(i{-}1),\alpha(i)}}\sum_{j=1}^{i{-}1}\frac{s_{\alpha(j),\alpha(i)}}{\sigma'_{\alpha(j),\alpha(i)}}.
\ee

This completes the proof of Proposition 3.

\bibliographystyle{JHEP}
\bibliography{ScatteringEquations}
\end{document}